\documentclass[showpacs,aps]{revtex4}
\usepackage{amssymb}
\usepackage{amsmath}
\usepackage{graphicx}
\usepackage{lscape}
\usepackage{booktabs}
\begin{document}

\title{The energy dependence of antiparticle to particle ratios in high energy $pp$ collisions{} }

\author{Wang Jiang-ling$^1$ Chen Gang$^{1,2}$\footnote{Email:chengang1@cug.edu.cn}\ \  Li Hai-jun$^1$ and Li Di-kai$^1$}

\address{$^1$School of Mathematics and Physics, China University of Geosciences, Wuhan
430074, China.\\
$^2$Key Laboratory of Quark and Lepton Physics
(MOE), Central China Normal University, Wuhan 430079,China\\
chengang1@cug.edu.cn}

\begin{abstract}

The energy dependence of antiparticle to particle ratios in $pp$
collisions of high energy is studied using the PACIEA and DCPC
model. The yield ratios of antimatter and matter for different
masses are measured at various energies. It is found that the yield
ratios of antimatter and matter increase with the increase of the
c.m energy of $pp$ collisions until they gradually approach to 1
after the c.m energy is more than 200~GeV. The distribution of
transverse momentum also has significant dependence on the energy
and mass, i.e the average transverse momentum increase when the c.m
energy of $pp$ collisions increase. The model results are compatible
with the STAR preliminary datum.
\end{abstract}



 \maketitle

\section{Instructions}

The investigation of antimatter has great importance in nuclear and
particle physics, astrophysics, and cosmology~\cite{yan1}. Matter
and antimatter existed in equal abundance during the initial stage
of the universe. However, after a series of evolution, now the
amount of antimatter is far less than the amount of matter. Since
the evolution of the initial fireball created in relativistic
heavy-ion collisions is similar to the initial stage of the
universe, the relativistic heavy-ion collision experiment is the
appropriate means of antimatter investigation.

In 1928, based on the application of symmetry rules of quantum
mechanics, the British physicist Dirac predicted that each particle
had a corresponding antiparticle, which has opened the prologue of
antimatter research~\cite{Dirac}. Then, in 1932, American scientists
Anderson found positrons in the cosmic rays, which caused a stir in
the scientific community~\cite{Anderson}, in the United States in
1955, Segre and Chamberlain found antiprotons in Berkeley proton
synchrotron experiments~\cite{Chamberlain}, then in 1956, Cork B. et
al. found the anti-neutron when accelerating the protons to 6.2 GeV
in Tevatron accelerator~\cite{Cork}. The discovery of antimatter on
the scale of nucleus, was started with the measurement of
anti-hydrogen in 1995, and finally in January 17, 2010, the
scientists captured 38 anti-hypertritium in LHC, CERN~\cite{Alpha}.

Recently, the STAR collaboration has reported their results in
measuring the Au-Au collisions at top RHIC energy region. They found
70$\pm$17 anti-tritium and 157$\pm$30 tritium in Au-Au collisions at
$\sqrt{s}$ =200 GeV, with 89$\times 10^6$ events and 22$\times 10^6$
central collision events. ALICE collaboration also published the
yield of anti-deuterium is approximately 6$\times 10^{-5}$ , in 7
TeV $pp$ collisions~\cite{Sharma1,Sharma2}. On the other hand, the
theoretically research is generally divided into two steps. First,
select a model to calculate the nucleons and hyperons, such as the
transport model. Then, use the phase-space coalescence
model~\cite{Mattiello} or statistical model~\cite{pop}to calculate
the light (anti-)nuclei. Some scientists use the
coalescence+blast-wave method ~\cite{L} and UrQMD hydro hybrid
model+thermal model~\cite{Steinheimer}to do the theoretical research
on the production of light nuclei and hyper nuclei in Au-Au and
Pb-Pb collisions at relativistic energy.

Based on PACIAE~\cite{yan2}, we proposed a new model, dynamically
constrained phase-space coalescence model (DCPC), to study light
(anti-)nuclear. Using this model, we have investigated the
production of light(anti)nuclei in relativistic $pp$
collisions~\cite{yan1}, the production of light (anti)nuclei and
(anti)hypertritium in Au-Au collisions~\cite{chen1}, and their
centrality dependence~\cite{chen2} and scaling feature~\cite{chen3}.
The consistency of the results of the model and the corresponding
experimental data shows that PACIAE+DCPC model is an effective
method which can be used to describe the production of light
(anti)nuclei and (anti)hypernucleus in relativistic heavy ion
collisions~\cite{yan1,chen2,chen3}.

In this paper, we firstly calculate the hadronic final state using
the PYTHIA~\cite{T} and PACIAE model in the non-single diffractive
(NSD) $pp$ collisions at different c.m energies. Then we can
generate the light(anti)nuclei using the dynamically constrained
phase-space model to investigate the energy dependence of
antiparticles production in high energy $pp$ collisions. In section
2, we briefly describe our model. In section 3, our numerical
results are presented and in section 4, we give a short summary.

\section{MODELS}
PYTHIA is a model for high energy hadron-hadron ($hh$)
collisions~\cite{T}. The parton and hadron cascade model,
PACIAE~\cite{yan2}, is based on PYTHIA. In the PYTHIA model a hh
collision is decomposed into parton-parton collisions. The hard
parton-parton collision is described by the lowest leading order
perturbative QCD (Lo-pQCD) parton-parton interactions with the
modification of parton distribution function in a hadron. The soft
parton-parton collision, which is non-perturbative phenomenon, is
considered empirically. The initial- and final-state QCD radiations
and the multiparton interactions are also taken into account. So the
consequence of a $hh$ collision is a partonic multijet state
composed of di-quarks (anti-diquarks), quarks (antiquarks), and
gluons, besides a few hadronic remnants. It is then followed by the
string construction and fragmentation and a hadronic final state for
a $hh$ (pp) collision is eventually obtained.

The PACIAE model is mainly different from the PYTHIA as follows:
\begin{enumerate}
\item The string fragmentation is switched-off temporarily and the di-quarks
(anti-diquarks) are broken randomly into quarks (antiquarks). So the
consequence of pp collision is a initial state of quarks,
antiquarks, and gluons, besides a few hadronic remnants. This
partonic initial state is regarded as the quark-gluon matter (QGM)
formed in the relativistic pp collisions.

\item The parton rescattering is introduced. In this stage the rescattering
among partons in QGM is calculated by the 2 $\rightarrow$ 2 Lo-pQCD
parton-parton interaction cross sections \cite{B}. However, a $K$
factor is introduced to include the higher order and
non-perturbative corrections. The effective strong coupling
constant, $\alpha_s$, is assumed to be 0.47. A parton colour screen
mass $\mu$=0.63 GeV is introduced to avoid the divergence. By
integrating the differential cross sections above the total cross
section of the parton collision is obtained. Then the parton
rescattering is simulated by Monte Carlo method.

\item The hadronization is then proceeded after parton rescattering. The
partonic matter can be hadronized by the Lund string fragmentation
regime~\cite{T} and/or the phenomenological coalescence model~\cite{yan2}.
\item At last the hadron rescattering is added. In this stage the hadronic
matter after hadronization proceeds rescattering. This rescattering
is dealt with by the usual two-body collision method~\cite{sa1},
until the hadronic freeze-out is reached(the exhaustion of
parton-parton collision pairs).
\end{enumerate}

From quantum statistical mechanics~\cite{K}, one can not precisely define
both position $\vec q\equiv (x,y,z)$ and momentum $\vec p \equiv
(p_x,p_y,p_z)$ of a particle in six dimension phase space, because
of the uncertainty principle,
$\Delta\vec q\Delta\vec p\geq h^3$.
One can only say this particle lies somewhere within a six dimension
quantum``box" or ``state" of volume of $\Delta\vec q\Delta\vec p$. A
$h^3$ volume element in the six dimension phase space corresponds to
a state of the particle~\cite{K}. Therefore, one can estimate the yield
of a single particle by
\begin{equation}
Y_1=\int_{H\leq E} \frac{d\vec qd\vec p}{h^3}.
\end{equation}
Similarly, the yield of N particles cluster can be estimated by
\begin{equation}
Y_N=\int ...\int_{H\leq E} \frac{d\vec q_1d\vec p_1...d\vec q_Nd\vec
p_N}{h^{3N}}. \label{phas}
\end{equation}

In the dynamically constrained phase space coalescence model, the
yield of $\overline{_{\overline\Lambda}^3H}$, for instance, is
assumed to be
\begin{align}
Y_{\overline{_{\overline\Lambda}^3H}}=&\int ...
\int\delta_{123}\frac{d\vec q_1d\vec p_1
  d\vec q_2d\vec p_2d\vec q_3d\vec p_3}{h^{9}},
\label{yield} \\
\delta_{123}=&\left\{
  \begin{array}{ll}
  1 \hspace{0.2cm} \textrm{if} \hspace{0.2cm} 1\equiv \bar p, 2\equiv \bar n,
    3\equiv \bar\Lambda, \textrm{and combination};\\
    \hspace{0.95cm} m_0\leqslant m_{inv}\leqslant m_0+\Delta m;\\
    \hspace{0.95cm} |\vec q_{12}|\leqslant D_0, \hspace{0.2cm}|\vec q_{13}|
    \leqslant D_0, \hspace{0.2cm}|\vec q_{23}|\leqslant D_0; \\
  0 \hspace{0.2cm}\textrm{otherwise},
  \end{array}
  \right.
\label{yield1}
\end{align}
where
\begin{equation}
m_{inv}=[(E_1+E_2+E_3)^2-(\vec p_1+\vec p_2+\vec p_3)^2]^{1/2},
\label{yield2}
\end{equation}
$m_0$ and $D_0$ stand for the rest mass and diameter of
$\overline{_{\overline\Lambda}^3H}$, $\Delta m$ refers to the
allowed uncertainty, and $|\vec q_{ij}|$ is the distance between
particles $i$ and $j$.

The integral over continuous distributions in Eq.~(\ref{yield})
should be replaced by the sum over discrete distributions as the
hadron position and momentum distributions from transport model
simulation are discrete. In a single event of the final hadronic
state from transport model simulation, the system of
$\overline{_{\overline\Lambda}^3H}$ can be identified by the
configurations of $\bar p$, $\bar n$, and $\bar\Lambda$. This
configuration can be expressed as
\begin{equation}
C_{\bar p\bar n\bar\Lambda}(\Delta q_1,\Delta q_2,\Delta q_3;\vec
p_1,\vec p_2,\vec p_3), \label{conf}
\end{equation}
where the subscripts 1,2,3 stand for the $\bar p$, $\bar n$ and
$\bar \Lambda$, and $\Delta q_i=|\vec q_i-\vec q_c| \ \ (i=1,2,3)$
refers to the distance between $i$th particle and the center-of-mass
of $\bar p$, $\bar n$, and $\bar\Lambda$ for instance.
Correspondingly, the third constraint (diameter constraint) in
Eq.~(\ref{yield1}) is replaced by
\begin{equation}
\Delta q_i\leqslant R_0, \ \ \ (i=1,2,3),
\end{equation}
where $R_0$ refers to the radius of
$\overline{_{\overline\Lambda}^3H}$.

Each configuration above contributes an partial yield of
\begin{equation}
Y_{123}=\left\{
  \begin{array}{ll}
  1 \hspace{0.2cm} \textrm{if} \hspace{0.2cm} m_0\leqslant m_{inv}\leqslant m_0+\Delta m,\\
    \hspace{0.4cm} q_1\leqslant R_0, \hspace{0.2cm} q_2\leqslant R_0, \hspace{0.2cm}
                   q_3\leqslant R_0; \\
  0 \hspace{0.2cm}\textrm{otherwise};
  \end{array}
  \right.
\label{yield3}
\end{equation}
to the $\overline{_{\overline\Lambda}^3H}$. Its total yield in a
single event is then the sum of above partial yield over the
configurations of Eq.~(\ref{conf}) and its combination. An average
over events is required at the end.


\section{The results}

In the PYTHIA and PACIAE simulations, we assume that the hyperons
heavier than $\Lambda$ decay already. The model parameters were
fixed on the default values given in PYTHIA model, except the $K$
factor and the parameters of parj(1), parj(2), and parj(3) were
roughly fitted to the STAR data of $\Xi^+, \Xi^-, \Lambda$, and
$\overline \Lambda$ in NSD $pp$ collisions at $\sqrt{s}$= 200
GeV~\cite{yan1}. Here the $K$ factor is introduced to consider the
higher order and non-perturbative QCD corrections. parj(1) is the
suppression of diquark-antidiquark pair production compared with
quark-antiquark production. parj(2) is the suppression of $s$ quark
pair production compared with $u$ ($d$) pair production. parj(3) is
the extra suppression of $s$ diquark production compared with the
normal suppression of $s$ quarks. The fitted parameters of $K$=3
(default value is 1 or 1.5), parj(1)=0.15 (0.1), parj(2)=0.38 (0.3),
and parj(3)=0.45 (0.4). We use the PYTHIA and PACIAE transport model
to generate the hadronic final states, and use the DPCP model to
generate the light (anti)nuclei, respectively, then to study the
characteristics of them.

In Figure 1, the yield ratios of antiparticles to corresponding
particles in $pp$ collisions at 200 GeV and 7 TeV are given,
including $\pi^-/\pi^+$, $K^-/K^+$, $\bar{p}/p$,
$\bar{\Lambda}/\Lambda$, ${\overline{\Xi^-}}/\Xi^-$,
${\overline{\Omega^-}}/\Omega^-$, $\bar{d}/d,
{\overline{^3He}}/^3He$, and ${\overline{_{\overline\Lambda}^3H}}
/{_{\overline\Lambda}^3H}$, respectively. For comparison,the
experimental results are also drawn in the Figure 1, as the dots
representing STAR data~\cite{BI} and the solid triangle is ALICE
data~\cite{KA}. From the figure 1, it can be seen that, all ratios
of antiparticles to particles are approximately close to 1 and most
of them is slightly less than 1, and the results of PYTHIA and
PACIAE Model have good repeatability. The results obtained from our
model are in agreement with the experimental data from
STAR~\cite{BI} and ALICE ~\cite{KA}.

\begin{figure}[th]
\includegraphics[width=10 cm]{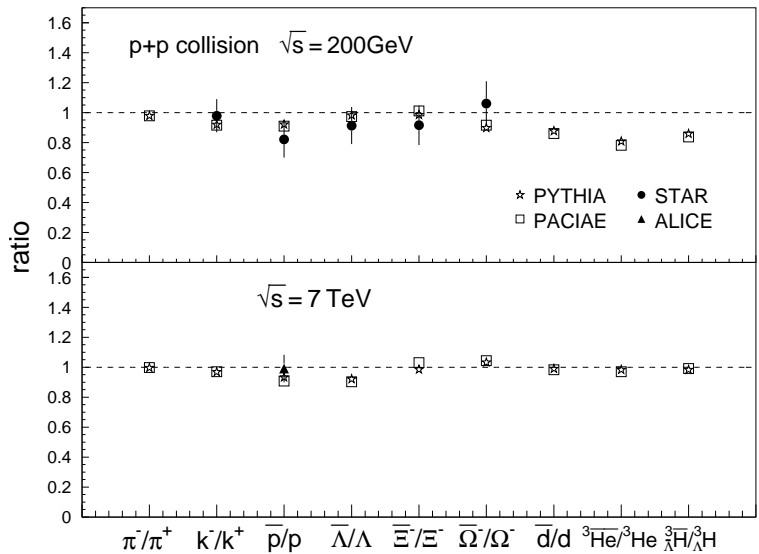}
\caption{The yield ratios of antiparticles ($\pi^-$, $K^-$,
$\bar{p}$, $\bar{\Lambda}$, ${\overline{\Xi^-}}$,
${\overline{\Omega^-}}$, $\bar{d}$, ${\overline{^3He}}$, and
${\overline{_{\overline\Lambda}^3H}}$) to corresponding particles
($\pi^+$, $K^+$, $p$, $\Lambda$, $\Xi^-$, $\Omega^-$, $d$, $^3He$,
and ${_{\overline\Lambda}^3H}$) in $pp$ collisions of 200 GeV and 7
TeV, where mesons and Baryons produced by PYTHIA and PACIAE Models,
the light (anti)nuclei and (anti)hypertritons produced by DCPC
model. The open symbols represent our model results. The solid
symbols are the data points from STAR{$~^{21}$} and
ALICE{$~^{22}$}.} \label{}

\end{figure}

In order to study the energy dependence of the yield ratio of
antiparticles to particles, the yield ratios of antiparticles to
particles are calculated by the PACIAE and DCPC model for different
center-of-mass (c.m) energies $pp$ collisions of 10~GeV, 17.3~GeV,
50~GeV, 100~GeV, 200~GeV, 1~TeV, 7~TeV, 14~TeV, respectively, as
shown in Figure 2. The ratios for the $\bar p/p$ and $\bar \Lambda
/\Lambda$ are obtained by PACIAE model, the ratios for the
$\bar{d}/d, {\overline{^3He}}/^3He$, and
${\overline{_{\overline\Lambda}^3H}} /{_{\overline\Lambda}^3H}$  are
calculated by PYTHIA + DCPC model for all energies.

We can see from the Figure 2 that the yield ratios ascend rapidly
with the rise of c.m energy. When c.m energy reaches beyond 200 GeV,
the yield ratios tend to saturate, and then the value of yield
ratios of all particles gradually tend to 1 as the c.m energy is
more than 1.0~TeV. It can also be noticed that the larger the mass
of the particle is, the smaller the ratio is. In low energy,
collisions frequency is relatively small, and the production cross
section of final state particles coalesced by initial
quark-antiquark pairs is also small, especially the cross section of
antiparticles.
When the beam energy of the collisions is increased, the system
created becomes almost net-baryon free~\cite{KA}. It means that
probability of antiparticles production will reach that of
particles. This energy dependence of the yield ratio of
antiparticles to particles in high $pp$ collisions is similar to the
relativistic heavy ion collisions~\cite{KA,A}.

\begin{figure}[th]
\includegraphics[width=10 cm]{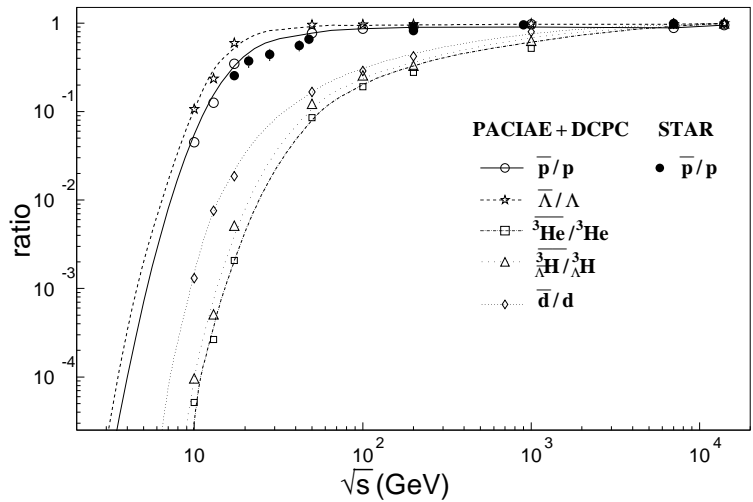}
\caption{The distribution of yield ratios of antiparticles ($\pi^-$,
$K^-$, $\bar{p}$, $\bar{\Lambda}$, ${\overline{\Xi^-}}$,
${\overline{\Omega^-}}$, $\bar{d}$, ${\overline{^3He}}$, and
${\overline{_{\overline\Lambda}^3H}}$) to corresponding particles
($\pi^+$, $K^+$, $p$, $\Lambda$, $\Xi^-$, $\Omega^-$, $d$, $^3He$,
and ${_{\overline\Lambda}^3H}$) in $pp$ collisions , as a function
of c.m energy, which the data are simulated by PACIAE + DCPC model.
The c.m energies are 10~GeV, 13~GeV, 17.3~GeV, 50~GeV, 100~GeV,
200~GeV, 1~TeV, 7~TeV,and 14~TeV respectively.}\label{}
\end{figure}

The transverse momentum distributions of (anti)particles produced by
PACIAE model for the different c.m. energies in $pp$ collisions are
investigated, containing Pions, Kaons, Protons and Lambdas,
respectively. The results at $\sqrt s$ = 17.3~GeV, 1000~GeV and
14000~GeV are shown in figure 3. From the figure 3, one can observe
that the transverse momentum distributions increase sharply from
$p_T =0$~GeV/C until a maximum peak around $p_T=0.2\sim$0.5~GeV/C is
reached. For one particle, the peak value of transverse momentum
distributions reduces with the c.m. energy increasing and
distribution is broadened. The transverse momentum distributions of
different particle, shown in Figures 3(a)-(d), have a similar
behavior, which the peak moves rightward with the increasing of mass
of particle. However, the transverse momentum distributions of
antiparticles have a similar behavior with particles, whereas the
peak of antiparticle distributions moves leftward and the becomes
narrower compared to the particles distributions.

\begin{figure}[th]
\includegraphics[width=10 cm]{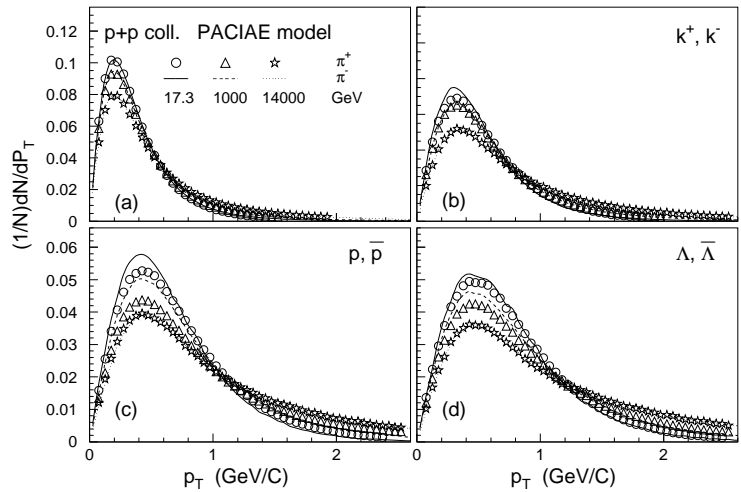}
\caption{The transverse momentum distributions of particles and
anti-particles produced by PACIAE model at $\sqrt s$ = 17.3~GeV,
1000~GeV and 14000~GeV in $pp$ collisions, respectively, as a
function of c.m energy. (a) $\pi^+ ,\pi^-$, (b) $K^+, K^-$, (c) $p,
\bar p$, (d) $\Lambda, \bar{\Lambda}$. } \label{}
\end{figure}

Figure 4 shows the energy dependence of average transverse momentum
of different particles, including anti-particles correspondingly. It
can be seen from the figure that the average transverse momentum
increases with the rise of energy. It can be seen from figure 3 that
the amount of particle in high $p_T$ region would increase with
higher energy. So we suggest that the rise of energy leads to more
violent collisions, larger transferring of momentum and energy, and
then larger cross section of hard process and hard particle
production, which obviously would raise the average transverse
momentum. Besides, we notice that the average transverse momentum of
mesons is smaller than that of baryons. It can be seen from figure 3
that the larger the mass of the particle is, the more the amount of
particle in high $p_T$ region is, then the larger the average
transverse momentum is. Since the mass of mesons is less than the
mass of baryons, the corresponding average transverse momentum of
mesons is smaller than that of baryon. Actually, the average
transverse momentum of particles is mass dependent, i.e. the
transverse momentum increases with the increasing of mass, this
conclusion is just consistent with the experimental
results~\cite{BI}.

\begin{figure}[th]
\includegraphics[width=10 cm]{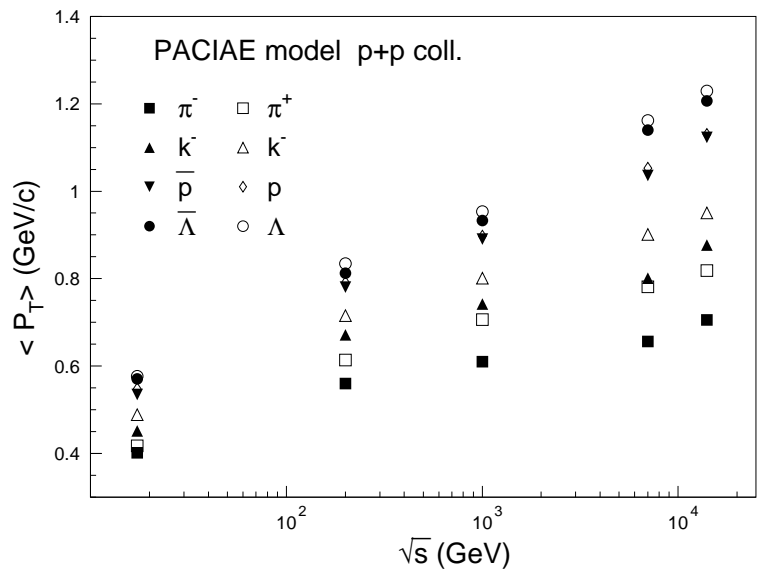}
\caption{The values of average transverse momentum for different
particles ($\pi^-$, $K^-$, $\bar{p}$, $\bar{\Lambda}$, $\pi^+$,
$K^+$, $p$, $\Lambda$) in $pp$ collisions, as a function of c.m
energy, which the data are simulated by PACIAE model. The c.m
energies are 17.3~GeV,  200~GeV, 1~TeV, 7~TeV,and 14~TeV
respectively.} \label{}
\end{figure}

\section{Conclusion}
We use PACIAE and PYTHIA model to generate final state hadrons of
different c.m energies in pp collisions and combine the light
(anti-)nuclei with the DCPC model. Firstly, the energy dependence of
antiparticle ($\pi^-$, $K^-$, $\bar{p}$, $\bar{\Lambda}$,
${\overline{\Xi^-}}$, ${\overline{\Omega^-}}$, $\bar{d}$,
${\overline{^3He}}$, and ${\overline{_{\overline\Lambda}^3H}}$) to
particle ($\pi^+$, $K^+$, $p$, $\Lambda$, $\Xi^-$, $\Omega^-$, $d$,
$^3He$, and ${_{\overline\Lambda}^3H}$) ratios in high energy $pp$
collisions is studied. The results show that antiparticle to
particle ratios increase with the increase of the c.m energy of $pp$
collisions from $\sqrt s=$ 10~GeV, until they gradually approach to
1 after the c.m energy is more than 200~GeV. Then the distribution
of transverse momentum of antiparticles and particles at different
c.m energies are investigated. It is found that the transverse
momentum distributions of different particles at different c.m
energies have a similar behavior, for one particle its peak reduce
and the distribution broaden with the increasing of the c.m energy
of particle. However, its peak moves rightward with the increasing
of mass of particle at same c.m energy. Contrary, the average
transverse momentum increase when the c.m energy of $pp$ collisions
increase. The model results are compatible with the STAR preliminary
datum.

\begin{center} {ACKNOWLEDGMENT} \end{center}
Finally, we acknowledge the financial support from NSFC
(11475149,11305144, 11303023) in China.

\end{document}